\documentclass[12pt]{article}
\newtheorem{theorem}{{Theorem}}
\newcommand{\bt}{\begin{theorem}}
\newcommand{\et}{\end{theorem}}

\newcommand{\newsection}[1]{\setcounter{equation}{0} \setcounter{theorem}{0}
\section{#1}}

\newcommand{\NI}{\noindent}
\newcommand{\bea}{\begin{eqnarray}}
\newcommand{\eea}{\end{eqnarray}}
\newcommand{\dsp}{\displaystyle}
\def \spec#1 {\mathop{#1}}

\def \b #1 {\bf #1}

\newcommand {\lgl}{\langle}
\newcommand {\rgl}{\rangle}

\newcommand {\CC}{\centerline}
\newcommand{\IC}{\hbox{\hspace{.20pc}{\sf I}\hspace{-.45pc}{\bf C}}}

\newcommand{\IP}{I\!\!P}

\newcommand{\cle}{{\cal E}}

\newcommand{\clh}{{\cal H}}

\newcommand{\clk}{{\cal K}}

\newcommand{\clp}{{\cal P}}

\newcommand{\cls}{{\cal S}}

\newcommand{\clv}{{\cal V}}

\def \bba {\mbox{\boldmath $a$}}
\def \bbb {\mbox{\boldmath $b$}}

\def \bbx {\mbox{\boldmath $x$}}
\def \bby {\mbox{\boldmath $y$}}

\newcommand{\sgm}{\sigma}

\newcommand{\ot}{\otimes}
\newcommand{\dg}{\dagger}

\newcommand{\op}{\oplus}
\newcommand{\lmd}{\lambda}

\newcommand{\al}{\alpha}

\newcommand{\raro}{\rightarrow}

\newcommand{\sbs}{\subset}

\newcommand{\vsp}{\vskip 1em}
\newcommand{\vspp}{\vskip 2em}

\newcommand{\ol}{\overline}

\def \qed {\hfill \vrule height6pt width 6pt depth 0pt}

\newcommand{\be}{\begin{equation}}
\newcommand{\ee}{\end{equation}}
\newcommand{\ben}{\begin{eqnarray*}}
\newcommand{\een}{\end{eqnarray*}}

\begin{document}
\CC {\large {\bf On the maximal dimension of a completely entangled }}
\CC {\large {\bf subspace for finite level quantum systems}}
\CC {by}
\CC {K.R. Parthasarathy}
\CC {Indian Statistical Institute, Delhi Centre}
\CC {7, S.J.S. Sansanwal Marg}
\CC {New Delhi 110 016, India}
\CC {e-mail : krp@isid.ac.in}

\vsp
\NI {\bf Summary :}   Let $\clh _i$  be a finite dimensional complex Hilbert space of dimension $d_i$ associated with a finite level quantum system $A_i$  for $i = i, 1,2, \ldots , k$. A subspace $S \sbs  \clh = \clh _{A_{1} A_{2}\ldots A_{k}} = \clh _1 \otimes \clh _2 \otimes \ldots \otimes \clh _k $
is said to be {\it completely entangled} if it has no nonzero product vector of the form $u_1 \otimes u_2 \otimes \ldots \otimes u_k$ with $u_i$ in $\clh _i$ for each $i$.  Using the methods of elementary linear algebra and the intersection theorem for projective varieties in basic algebraic geometry we prove that 
$$\max _{S \in \cle } \dim S = d_1 d_2\ldots d_k - (d_1 + \cdots + d_k) + k - 1$$
where $\cle $ is the collection of all completely entangled subspaces. 

When $\clh _1 = \clh _2 $ and $k = 2$ an explicit orthonormal basis of a maximal completely entangled subspace of $\clh _1 \otimes \clh _2$  is given. 

We also introduce a more delicate notion of a {\it perfectly entangled} subspace for a multipartite quantum system, construct an example using the theory of stabilizer quantum codes and pose a problem.
\vsp
\NI {\bf Key Words :}  finite level quantum systems, separable states, entangled states, completely entangled subspaces, perfectly entangled subspace, stabilizer quantum code. 
\vsp
\NI {\bf MSC index :}  81P68,  94B99
\vspp
\newsection  {Completely Entangled Subspaces}

Let $\clh _i$ be a complex finite dimensional Hilbert space of dimension $d_i$ associated with a finite level quantum system $A_i$ for each $i = 1,2,\ldots , k$.  A state $\rho $ of the combined system $A_i A_2 \ldots A_k$ in the Hilbert space
\be
\clh = \clh \otimes \clh _2 \otimes \ldots \ot \clh _k
\ee
is said to be  {\it separable} if it can be expressed as
\be
\rho = \dsp {\sum ^m_{i = 1}} p_i \rho _{i1} \ot \rho _{i2} \ot \ldots \ot \rho _{ik}
\ee
where $\rho _{ij}$ is a state of $A_j$ for each $j, p_i > 0$ for each $i$ and $\dsp {\sum ^m_{i=1}}p_i = 1$ for some finite $m$.  A state which is not separable is said to be {\it entangled}.  Entangled states play an important role in quantum teleportation and communication [3].  The following theorem due to Horodecki et al [2] suggests a method of constructing entangled states. 
\vsp
\NI {\bf Theorem 1.1}  (Horodecki et al)  Let $\rho $ be a separable state in $\clh $.  Then the range of $\rho $ is spanned by a set of product vectors.

For the sake of readers' convenience and completeness we furnish a quick proof. 
\vspp
\NI {\bf Proof :}  Let $\rho $ be of the form (1.2).  By spectrally resolving each $\rho _{ij}$ into one dimensional projections we can rewrite (1.2) as
\be
\rho = \dsp {\sum ^n_{i=1}} q_i | u_{i1} \ot u_{i2} \ot \ldots \ot u_{ik} \rgl \lgl u_{i1} \ot u_{i2} \ot \ldots \ot u_{ik} |
\ee
where $u_{ij}$ is a unit vector in $\clh _j$ for each $i,j$ and $q_i > 0$ for each $i$ with $\dsp {\sum ^n_{i=1}} q_i = 1$.  We shall prove the theorem by showing that each of the product vectors $u _{i1} \ot u_{i2} \ot \ldots \ot u_{ik}$ is, indeed, in the range of $\rho $.  Without loss of generality, consider the case $i=1$.  Write (1.3) as
\be
\rho = q_1 |u_{11} \ot u_{12} \ot \ldots \ot u_{1k}\rgl  \lgl u_{11} \ot u_{12} \ot \ldots \ot u_{1k}| + T 
\ee
where $q_1 > 0$ and $T$ is a nonnegative operator.  Suppose $\psi \neq 0$ is a vector in $\clh $ such that $T|\psi \rgl = 0$ and $\lgl u_{11} \ot u_{12} \ot \ldots \ot u_{1k}|\psi \rgl \neq 0$.  Then $\rho | \psi \rgl $ is a nonzero multiple of the product vector $u_{11} \ot u_{12} \ot \ldots \ot u_{1k}$ and $u_{11} \ot u_{12} \ot \ldots \ot u_{1k} \in R(\rho )$, the range of $\rho $.  Now suppose that the null space $N(T)$ of $T$ is contained in $\{ u_{11} \ot u_{12} \ot \ldots \ot u_{1k}\}^\perp $.  Then $R(T) \supset \{ u_{11} \ot u_{12} \ot \ldots \ot u_{1k}\} $ and therefore there exists a vector $\psi \neq 0$ such that 
$$T | \psi \rgl = |u_{11} \ot u_{12} \ot \ldots \ot u_{1k} \rgl . $$
Note that $\rho | \psi \rgl \neq 0$, for otherwise, the positivity of $\rho , T$ and $q_1$ in (1.4) would imply $T | \psi \rgl = 0$.  Thus (1.4) implies
$$\rho |\psi \rgl = (q_1 \lgl u_{11} \ot u_{12} \ot \ldots \ot u_{1k} | \psi \rgl + 1 )~|u_{11}\ot u_{12} \ot \ldots \ot u_{1k} \rgl . ~~~~~~~~~~\qed $$
\vsp
\NI {\bf Corollary }  If a subspace $S \sbs \clh _1 \ot \clh _2 \ot \ldots \ot \clh _k $ does not contain any nonzero product vector of the form $u_1 \ot u_2 \ot \ldots \ot u_k$ where $u_i \in \clh _i$ for each $i$, then any state with support in $S$ is entangled. 
\vspp
\NI {\bf Proof :}  Immediate.   \qed
\vspp
\NI {\bf Definition 1.2}   A nonzero subspace $S \sbs \clh $ is said to be {\it completely entangled} if $S$ contains no nonzero product vector of the form $u_1 \ot u_2 \ot \ldots \ot u_k$ with $u_i \in \clh _i$ for each $i$. 

Denote by $\cle $ the collection of all completely entangled subspaces of $\clh $.  Our goal is to determine $\dsp {\max _{S\in \cle }} \dim S $. 
\vspp
\NI {\bf Proposition 1.3}  There exists $S \in \cle $ satisfying
$$\dim~S = d_1 d_2 \ldots d_k - (d_1 + d_2 + \cdots d_k) + k-1 . $$
\vsp
\NI {\bf Proof :}   Let $N = d_1 + d_2 + \cdots + d_k - k+1$.  Without loss of generality, assume that $\clh _i = \IC ^{d_{i}}$ for each $i$,  with the standard scalar product.  Choose and fix a set $\{ \lmd _1, \lmd _2, \ldots , \lmd _N\} \sbs \IC $ of cardinality $N$.  Define the colum vectors 
\be
u_{ij} = \left [ \begin{array} {c}1 \\ \lmd _i \\ \lmd ^2_i \\ \vdots \\ \lmd ^{d_{j}-1}_i \end{array} \right ] , 1 \le i \le N, ~~ 1 \le j \le k 
\ee
and consider the subspace
\be
S = \{ u_{i1} \ot u_{i2} \ot \ldots \ot u_{ik}, ~~1 \le i \le N\}^{\perp} \sbs \clh . 
\ee
We claim that $S$ has no nonzero product vector.  Indeed, let
$$0 \neq v_1 \ot v_2 \ot \ldots \ot v_k \in S, ~~ v_i \in \clh _i . $$
Then
\be
\dsp {\prod ^k_{j = 1}} \lgl v_j | u_{ij} \rgl = 0, ~~ 1 \le i \le N. 
\ee
If
\be
E_j = \{ i | \lgl v_j | u_{ij} \rgl = 0 \} ~ \sbs ~ \{ 1,2, \ldots , N\}
\ee
then (1.7) implies that 
$$ \{ 1,2, \ldots , N\} = \dsp {\cup ^k_{j=1}} E_j $$
and therefore 
$$N \le \sum ^k _{j=1}  \# E_j. $$
By the definition of $N$ it follows that for some $j, ~\# E_j \ge d_j$. Suppose $\# E_{j_{0}} \ge d_{j_{0}}$.  From (1.8) we have 
$$\lgl v_{j_{0}} | u_{ij_{0}}\rgl = 0~~\mbox {for}~~ i = i_1, i_2, \ldots , i_{d_{j_{0}}} $$
where $i _1 < i_2 < \cdots < i_{d_{j_{0}}} $.  From (1.5) and the property of van der Monde determinants it follows that $v_{j_{0}} = 0$, a contradiction.  Clearly, $\dim ~ S \ge d_1d_2 \cdots d_k - (d_1 + \cdots + d_k) + k-1. $ \qed 
\vspp
\NI {\bf Prposition 1.4 }  Let $S \sbs \clh $ be a subspace of dimension $d_1 d_2 \ldots d_k - (d_1 + \cdots + d_k)+ k$.  Then $S$ contains a nonzero product vector.
\vsp
\NI {\bf Proof  :}  Identify $\clh _j$ with $\IC ^{d_{j}}$ for each $j = 1,2,\ldots , k$.  For any nonzero element $v$ in a complex vector space $\clv $ denote by $[v]$ the equivalence class of $v$ in the projective space $\IP (\clv )$.  Consider the map
$$T : \IP (\IC^{d_{1}}) \times \IP (\IC ^{d_{2}})\times \cdots \times \IP (\IC ^{d_{k}}) \raro \IP (\IC ^{d_{1}} \ot \IC ^{d_{2}} \ot \ldots \ot \IC ^{d_{k}}) $$
given by 
$$T ([u_1], [u_2], \ldots , [u_k]) = [u_1 \ot \ldots \ot u_k ]. $$
The map $T$ is algebraic and hence its range $R(T)$ is a complex projective variety of dimension $\dsp {\sum ^k_{i=1}} (d_i - 1)$.  By hypothesis $\IP (S)$ is a projective variety of dimension $d_1 d_2 \ldots d_k - (d_1 + \ldots + d_k) + k-1$.  Thus
\ben
\dim \IP (S) + \dim R(T) & = & d_1 d_2 \ldots d_k - 1 \\
& = & \dim \IP (\IC ^{d_{1}} \ot \IC ^{d_{2}} \ot \ldots \ot \IC ^{d_{k}}).
\een
Hence by Theorem 6, page 76 in [4] we have
$$ \IP (S) \bigcap R(T) \neq \emptyset . $$
In other words $S$ contains a product vector.  \qed
\vspp
\NI {\bf Theorem 1.5}  Let $\cle $ be the collection of all completely entangled subspaces of $\clh _1 \ot \clh _2 \ot \ldots \ot \clh _k $.  Then 
$$\max _{S \in \cle } \dim S = d_1 d_2 \ldots d_k - (d_1 + d_2 + \cdots + d_k) + k-1. $$
\vsp
\NI {\bf Proof  :}  Immediate from Proposition 1.3 and Proposition 1.4.  \qed
\vspp
\newsection {An Explicit Orthonormal Basis for a Completely Entangled Subspace of Maximal Dimension in $\IC ^n \ot \IC ^n $}

Let $\{ |x \rgl , x =  0,1,2,\ldots , n-1\} $ be a labelled orthonormal basis in the Hilbert space $\IC ^n$.  Choose and fix a set
$$E = \{ \lmd _1, \lmd _2, \ldots , \lmd _{2n-1}\} \sbs \IC $$
of cardinality $2n-1$ and consider the subspace
$$ S= \{  u_{\lmd _{i}} \ot u_{\lmd _{i}} , 1 \le i \le 2 n-1\}^{\perp}$$
where
$$u_\lmd = \sum ^{n-1}_{x=0} \lmd ^x | x \rgl , \lmd \in ~ \IC . $$
By the proof of Proposition 1.3 and Theorem 1.5 it follows that $S$ is a maximal completely entangled subspace of dimension $n^2 - 2n + 1$.  We shall now present an explicit orthonormal basis for $S$.  

First, observe that $S$ is orthogonal to a set of symmetric vectors and therefore $S$ contains the antisymmetric tensor product space $\IC ^n \wedge \IC^n $ which has the orthonormal basis  
\be
B_0 = \left \{ \frac {|xy\rgl - | yx \rgl }{\sqrt {2}} , 0 \le x < y \le n-1 \right \} . 
\ee
Thus, in order to construct an orthonormal basis of $S$, it is sufficient to search for symmetric tensors lying in $S$ and constituting an orthonormal set.  Any symmetric tensor in $S$ can be expressed as
\be
\sum _{{0\le x \le n-1}\atop {0\le y \le n -1}} f(x,y) | xy\rgl
\ee
where $f(x,y) = f(y,x)$ and
$$\sum _{{0\le x \le n-1}\atop {0\le y \le n-1}} f(x,y) \lmd ^{x+y}_i = 0, ~~~ 1 \le i \le 2n-1, $$
which reduces to 
\be
\sum _{{0\le x \le n-1}\atop {0\le j -x \le n-1}} f (x,j - x) = 0  ~\forall ~ 0 \le j \le 2n - 2.
\ee
Define $\clk _j$ to be the subspace of all symmetric tensors of the form (2.2) where the coefficient function $f$ is symmetric, has its support in the set $\{ (x, j - x), 0 \le x \le n-1, 0\le j - x \le n-1\} $ and satisfies (2.3).  Simple algebra shows that $\clk _0 = \clk _1 = \clk _{2n-3} = \clk _{2n-2} = 0$ and
$$S = \clh \wedge \clh \op \dsp {\op ^{2n-4}_{j=2}} \clk _j . $$
We shall now present an orthonormal basis $B_j$ for $\clk _j , 2\le j \le 2n-4.$  This falls into four cases. 
\vsp
\NI {\bf Case 1 :}  $2\le j \le n-1,~~j$ even
$$B_j = \left \{ \frac {1}{\sqrt {j(j+1)}} \left [ \sum ^{\frac {j}{2}-1}_{m=0} (|m~j-m\rgl + |j-m~m\rgl ) - j \left |\frac {j}{2} ~\frac {j}{2}\right \rgl \right ] \right \} $$
$$\cup \left \{ \frac {1}{\sqrt {j}}\sum ^{\frac {j}{2}-1}_{m=0} e^{\frac {4i\pi mp}{j}} (|m~j-m\rgl + |j - m ~ m\rgl ), ~~1 \le p \le \frac {j}{2} -1\right \} . $$
\vsp
\NI {\bf Case 2  :} $2\le j \le n -1,~~ j$ odd

$$B_j = \left \{ \frac {1}{\sqrt {j+1}}\sum ^{\frac {j-1}{2}}_{m=0} e^{\frac {4i\pi mp}{j+1}} (|m~j-m\rgl + |j - m ~ m\rgl ), ~~1 \le p \le \frac {j-1}{2}\right \} . $$
\vsp
\NI {\bf Case 3  :} $n \le j \le 2n -4,~~j$ even
\ben
B_j & = & \left \{ \frac {1}{\sqrt {(2n-2-j)(2n-1-j)}}\left [\sum ^{\frac {2n-2-j}{2}-1}_{m=0} (|j-n+m+1~ n-m-1\rgl  \right . \right .\\
& & \left .\left .+ |n-m-1~j-n+m+1\rgl) - (2n-2-j)  | \frac {j}{2} \frac {j}{2}\rgl \right ] \right \} 
\een
\ben
\cup & & \left \{ \frac {1}{\sqrt {2n-2-j}}\sum ^{\frac {2n-2-j}{2}-1}_{m=0} e^{\frac {4i\pi mp}{2n-2-j}} (|j-n+m+1~n-m-1\rgl \right . \\
& & \left . |n-m-1~j-n+m+1\rgl ),1 \le p \le \frac {2n-2-j}{2} - 1 \right \} 
\een
\vsp
\NI {\bf Case 4  :}  $n \le j \le 2n -4,~~ j$ odd
\ben
B_j & = & \left \{ \frac {1}{\sqrt {2n-1-j}}\sum ^{\frac {2n-1-j}{2}-1}_{m=0} e^{\frac {4i \pi mp}{2n-1-j}} (|j-n+m+1~n-m-1\rgl \right . \\
& + & \left . |n-m-1~j-n+m+1\rgl ),1 \le p \le \frac {2n-1-j}{2} - 1 \right \} 
\een
\vsp
The set $B_0 \cup \dsp {\cup ^{2n-4}_{j=2}} B_j$, where $B_0$ is given by (2.1) and the remaining $B_j$'s are given by the four cases above constitute an orthonormal basis for the maximal completely entangled subspace $S$.  
\vspp
\newsection {Perfectly Entangled Subspaces}

As in Section 1, let $\clh _i$ be a complex Hilbert space of dimension $d_i$ associated with a finite level quantum system $A_i$ for each $i = 1,2,\ldots , k$.  For any subset $E \sbs \{ 1,2,\ldots , k\} $ let 
$$\clh (E) = \ot _{i \in E} \clh _i $$
$$d (E) = \prod _{i\in E} d_i $$
so that the Hilbert space $\clh = \clh (\{ 1,2,\ldots , k\} )$ of the joint system $A_1 A_2 \ldots A_k$ can be viewed as $\clh (E)\ot \clh (E'), E'$ being the complement of $E$.  For any operator $X$ on $\clh $ we write
$$X(E) = Tr_{\clh (E')} X$$
where the right hand side denotes the relative trace of $X$ taken over $\clh (E')$.  Then $X(E)$ is an operator in $\clh (E)$.  If $\rho $ is a state of the system $A_1 A_2 \ldots A_k$ then $\rho (E)$ describes the marginal state of the subsystem $A_{i_{1}}A_{i_{2}} \ldots A_{i_{r}}$ where $E = \{ i_1, i_2,\ldots , i_r\}. $
\vspp
\NI {\bf Definition 3.1}  A nonzero subspace $\cls \sbs \clh $ is said to be perfectly entangled if for any $E \sbs \{ 1,2,\ldots , k\} $ such that $d (E) \le d(E')$ and any unit vector $\psi \in \cls $ one has
$$(|\psi \rgl \lgl \psi |) ~(E) = \frac {I_E}{d(E)} $$
where $I_E$ denotes the identity operator in $\clh (E)$.

For any state $\rho ,$ denote by $S(\rho )$ the von Neumann entropy of $\rho $.  If $\psi $ is a pure state in $\clh $ then $S ((|\psi \rgl  \lgl \psi |)~(E)) = S((|\psi \rgl  \lgl \psi |)~(E')). $  Thus perfect entanglement of a subspace $\cls $ is equivalent to the property that for every unit vector $\psi $ in $\cls $, the pure state $|\psi \rgl \lgl \psi |$ is maximally entangled in every decomposition $\clh (E) \ot \clh (E')$, i.e.,
$$S((|\psi \rgl \lgl \psi |) (E)) = S((|\psi \rgl \lgl \psi |)(E')) = \log _2 d(E) $$
whenever $d(E) \le d(E')$.  In other words, the marginal states of $|\psi \rgl  \lgl \psi |$ in $\clh (E)$ and $\clh (E')$ have the maximum possible von Neumann entropy.

Denote by $\clp $ the class of all perfectly entangled subspaces of $\clh $.  It is an interesting problem to construct examples of perfectly entangled subspaces and also compute $\dsp {\max _{\cls \in \clp}}\dim \cls $. 

Note that a perfectly entangled subspace $\cls $ is also completely entangled.  Indeed, if $\cls $ has a unit product vector $\psi = u_1\ot u_2 \ot \cdots \ot u_k$ where each $u_i$ is a unit vector in $\clh _i$  then $(|\psi \rgl \lgl \psi |) (E)$ is also a pure product state with von Neumann entropy zero.  Perfect entanglement of $\cls $ implies the stronger property that every unit vector $\psi $ in $\cls $ is {\it indecomposable}, i.e., $\psi $ cannot be factorized as $\psi _1 \ot \psi _2$ where $\psi _1 \in \clh (E), \psi _2 \in \clh (E')$ for any proper subset $E \sbs \{ 1,2,\ldots , k\} $.  
\vsp
\NI {\bf Proposition 3.2}  Let $\cls \sbs \clh $ be a subspace and let $P$ denote the orthogonal projection on $\cls $.  Then $\cls $ is perfectly entangled if and only if, for any proper subset $E \sbs \{ 1,2,\ldots , k\}$ with $d(E)\le d (E')$
$$(P X P) (E) = \frac {Tr ~ PX}{d(E) } I_E $$
for all operators $X$ on $\clh $. 
\vsp
\NI {\bf Proof :} Sufficiency is immediate.  To prove necessity, assume that $S$ is perfectly entangled.  Let $X$ be any hermitian operator on $\clh $.  Then by spectral theorem and Definition 3.2 it follows that $(P X P) (E) = \al (X) I_E$ where $\al (X)$ is a scalar.  Equating the traces of both saides we see that $\al (X) = d(E)^{-1} Tr PX $.  If $X$ is arbitrary, then $X$ can be expressed as $X_1 + i X_2$ where $X_1$ and $X_2$ are hermitian and the required result is immediate.  \qed

Using the method of constructing single error correcting 5 qudit stabilizer quantum codes in the sense of Gottesman [1], [3] we shall now describe an example of a perfectly entangled $d$-dimensional subspace in $h^{\ot ^{5}}$ where $h$ is a $d$-dimensional Hilbert space.  To this end we identify $h$ with $L^2 (A)$ where $A$ is an abelian group of cardinality $d$ with group operation + and null element $0$.  Then $h^{\ot ^{5}}$ is identified with $L^2 (A^5)$.  For any $\bbx  = (x_0, x_1, x_2, x_3, x_4)$ in $A^5$ denote by $|\bbx \rgl $ the indicator function of the singleton subset $\{ \bbx \}$ in $A^5$.  Then $\{ |\bbx \rgl , \bbx \in A^5\} $ is an orthonormal basis for $h^{\ot ^{5}}$.  Choose and fix a nondegenerate symmetric bicharacter $\lgl .~,~.\rgl $ for the group $A$ satisfying the following :
$$|\lgl a, b\rgl | =1, \lgl a,b\rgl = \lgl b,a\rgl, \lgl a,b+c\rgl = \lgl a,b\rgl \lgl a,c\rgl ~\forall ~a,b,c \in A$$
and $a = 0$ if and only if $\lgl a,x \rgl = 1$ for all $x \in A$.  Define
$$\lgl \bbx ,~\bby \rgl = \dsp {\prod ^{4}_{i=0}} \lgl x_i, ~y_i\rgl , ~ \bbx ,~\bby  \in A^5.$$
(Note that $\lgl \bbx ,~\bby \rgl $ denotes the bicharacter evaluated at $\bbx ,~\bby $ whereas $\lgl \bbx  |\bby  \rgl $ denotes the scalar product in $\clh = L^2 (A^5).$) With these notations we introduce the unitary Weyl operators $U_{\bba }, V_{\bbb}$ in $\clh $ satisfying
$$U_{\bba }|\bbx \rgl  = |\bba  + \bbx \rgl , ~V_{\bbb}|\bbx \rgl = \lgl \underline {b},\bbx \rgl ~|\bbx \rgl ,~ \bbx  \in A^5. $$
Then we have the Weyl commutation relations:
$$U_{\bba }U_{\bbb} = U_{\bba  + \bbb},~ V_{\bba }V_{\bbb}= V_{\bba  + \bbb}, V_{\bbb}U_{\bba } = \lgl \bba , \bbb\rgl U_{\bba }V_{\bbb} $$
for all $\bba , \bbb \in A^5 $.  The family $\{ d^{-\frac{5}{2}}U_{\bba }V_{\bbb}, \bba , \bbb \in A^5 \} $ is an orthonormal basis for the Hilbert space of all operators $X,Y$ with the scalar product $\lgl X |Y\rgl = Tr X^\dg Y$ between two operators $X, Y$.

Introduce the cyclic permutation $\sgm $ in $A^5$ defined by 
\be
\sgm ((x_0, x_1, x_2, x_3, x_4)) = (x_4, x_0, x_1, x_2, x_3). 
\ee
Then $\sgm $ is an automorphism of the product group $A^5$ and
$$ \sgm ^{-1} ((x_0, x_1, x_2, x_3, x_4)) = (x_1, x_2, x_3, x_4, x_0). $$
Define
\be
\tau (\bbx ) = \sgm ^2 (\bbx ) + \sgm ^{-2} (\bbx ).
\ee
Let $C \sbs A^5$be the subgroup defined by
$$C = \{ \bbx |x_0 + x_1 + x_2 + x_3 + x_4 = 0\} . $$
Define
\be
W_{\bbx } = \lgl \bbx , \sgm ^2 (\bbx )\rgl U_{\bbx }V_{\tau (\bbx )}, \bbx  \in A^5.
\ee
Then the correspondence $\bbx  \raro W_{\bbx } $ is a unitary representation of the subgroup $C$ in $\clh $.  Define the operator $P_C$ by 
\be
P_C = d^{-4} \sum _{\bbx  \in C} W_{\bbx } .
\ee
Then $P_C$ is a projection satisfying $Tr P_C = d$.  The range of $P_C$ is an example of a stabilizer quantum code in the sense of Gottesman.  From the methods of [1] it is also known that $P_C$ is a single error correcting quantum code.  The range $R(P_C)$ of $C$ is given by 
$$R(P_C) = \{ |\psi \rgl | W_{\bbx }|\psi \rgl = |\psi \rgl ~~\mbox{for all}~~\bbx  \in C\} . $$
Our goal is to establish that $R(P_C)$ is perfectly entangled in $L^2 (A)^{\ot^{5}}$.  To this end we prove a couple of lemmas. 
\vsp
\NI {\bf Lemma 3.3}  For any $\bba , \bbb  \in A^5$ the following holds :
$$\lgl \bba  |P_C | \bbb \rgl = \left \{ \begin{array}{l} 0 ~~\mbox{if}~\sum ^4_{i=0} (a_i - b_i) \neq 0, \\
d^{-4} \lgl \bba , \sgm ^2(\bba )\rgl \lgl \ol{\bbb , \sgm^2(\bbb )}\rgl ~\mbox{otherwise}. \end{array} \right .  $$
\vsp
\NI {\bf Proof  :}  We have from (3.1) - (3.4)
$$\lgl \bba  | P_C | \bbb \rgl = d^{-4} \sum _{x_{0}+x_{1}+x_{2}+x_{3}+x_{4} = 0} \lgl \bbx  , \sgm ^2 (\bbx )\rgl \lgl \tau (\bbx ), \bbb \rgl  \lgl \bba  |\bbx  + \bbb  \rgl $$
which vanishes if $\dsp {\sum ^{4}_{i=0}} (a_i - b_i)\neq 0$.  Now assume that $\dsp {\sum ^{4}_{i=0}} (a_i - b_i) = 0$.  Then
\ben
\lgl \bba  | P_C|\bbb  \rgl & = & d^{-4} \lgl \bba -\bbb , \sgm ^2(\bba -\bbb )\rgl  \lgl \sgm ^2 (\bba  - \bbb ), \bbb \rgl  \lgl \bba  - \bbb , \sgm ^2 \bbb \rgl \\
& = & d^{-4} \lgl \bba , \sgm ^2 (\bba  \rgl  \lgl \ol{\bbb , \sgm ^2(\bbb )}\rgl . ~~~~~~~~~~~~~~~~~~~~~~~~~~~~~~~~\qed
\een 
\vsp
\NI {\bf Lemma 3.4}  Consider the tensor product Hilbert space
$$L^2 (A)^{\ot ^{5}} = \clh _0 \ot \clh _1 \ot \clh _2 \ot \clh _3 \ot \clh _4 $$
where $\clh _i$ is the $i$-th copy of $L^2(A)$.  Then for any $(\{i,j\}) \sbs \{ 0,1,2,3,4\} $ and $\bba , \bbb  \in A^5$ the oprator $(P_C |\bba  \rgl  \lgl \bbb  |P_C)~ (\{ i,j\})$ is a scalar multiple of the identity in $\clh _i \ot \clh _j$.  
\vsp
\NI {\bf Proof :}   By Lemma 3.2 and the definition of relative trace we have, for any $x_0, x_1, y_0, y_1 \in A $, 
\ben
\lefteqn {\lgl x_0, x_1|(P_C | \bba  \rgl  \lgl \bbb  | P_C) ~(\{ 0,1\} ) | y_0, y_1 \rgl } \\ 
& = & \sum _{x_{2}, x_{3},x_{4} \in A} \lgl x_0, x_1, x_2, x_3, x_4 | P_C |\bba \rgl \lgl \bbb |P_C|y_0, y_1,x_2,x_3,x_4\rgl \\
& = & d^{-8} \sum _{{x_2+x_3+x_4 = \sum a_i-x_0-x_1}\atop {x_2 + x_3+x_4 = \sum b_i - y_0 - y_1 }} \lgl \bbx ,\sgm ^2 (\bbx )\rgl  \lgl \ol {\bba ,\sgm ^2(\bba )}\rgl \lgl \bbb , \sgm ^2 (\bbb )\rgl \\
& \times & \lgl y_0,y_1,x_2,x_3,x_4,\sgm ^2(y_0, y_1, x_2,x_3,x_4)\rgl
\een
The right hand side vanishes if $\sum (a_i - b_i) \neq x_0 + x_1 - y_0 - y_1) $.   Now suppose that $\sum (a_i - b_i) = x_0 + x_1 - y_0 - y_1 $.  Then the right hand side is equal to
\ben
\lefteqn {d^{-8} \lgl \bba , \sgm ^2 (\bba )\rgl  \lgl \bbb  , \sgm ^2 (\bbb)\rgl  \lgl \sum a_i - x_0 - x_1, x_0 + x_1 - y_0 - y_1\rgl } \\
\lefteqn { \times \sum _{x_{2}, x_4 \in A}\lgl x_2, y_1 - x_1 \rgl  \lgl x_4, y_0 - x_0\rgl } \\
& = & \left \{ \begin{array} {cc} 0 & \mbox{if} ~~x_0 \neq y_0~~\mbox{or}~~x_1 \neq y_1 , \\
d^{-6} \lgl \bba , \sgm ^2 (\bba )\rgl  \lgl \bbb , \sgm ^2 ( {\bbb})\rgl & \mbox{otherwise} . \end{array} \right . 
\een
This proves the lemma when $i = 0, j = 1$.  A similar (but tedious) algebra shows that the lemma holds when $i = 0, j = 2$. 

The cyclic permutation $\sgm $ of the basis $\{ |\bbx \rgl , \bbx  \in A^5\} $ induces a unitary operator $U_\sgm $ in $A^5$.  Since $\sgm $ leaves $C$ invariant it follows that $U_\sgm P_C = P_C U_\sgm $ and therefore
$$U_\sgm P_C |\bba  \rgl  \lgl \bbb  |P_C U^{-1}_\sgm = P_C |\sgm (\bba ) \rgl  \lgl \sgm (\bbb )|P_C, $$
which, in turn, imples that
\ben
\lefteqn {\lgl x_1, x_2 | (P_C |\bba \rgl  \lgl \bbb |P_C ) ~ (\{ 1,2\})|y_1, y_2 \rgl } \\
& = & \lgl x_1, x_2 | P_C |\sgm ^{-1} (\bba ) \rgl  \lgl \sgm ^{-1} (\bbb )|P_C) (\{ 0,1\})|y_1, y_2 \rgl . 
\een
By what has been already proved the lemma follows for $i = 1, j=2$.  A similar covariance argument proves the lemma for all pairs $\{ i,j\} .$  \qed
\vsp
\NI {\bf Theorem 3.4}  The range of $P_C$ is a perfectly entangled subspace of $L^2 (A)^{\ot ^{5}}$ and $\dim P_C = \# A.$
\vsp
\NI {\bf Proof  :}  Immediate from Lemma 3.3 and the fact that every operator in $L^2 (A^{\ot ^{5}})$ is a linear combination of operators of the form $|\bba \rgl  \lgl \bbb  |$ as $\bba , \bbb  $ vary in $A^5$.  \qed
\vsp
\NI {\bf Acknowledgement  :}  I thank A.J.Parameswaran and B. Pati for their help in explaining to me the role of basic algebraic geometry in the proof of Proposition 1.4.

\vspp

\CC {\bf REFERENCES}

\begin{enumerate}

\item  V. Arvind and K.R. Parthasarathy : A family of quantum stabilizer codes based on the Weyl commutation relations over a finite filed, in A Tribute to C.S. Seshadri, Perspectives in Geometry and Representation Theory, Edited by V. Lakshmibai et al, Hindustan Book Agency (India) 2003, pp 133 - 153.

\item M. Horodecki and R. Horodecki, Separability of mixed states : necessary and sufficient conditions, Phys. Lett.  A 223 (1-2) pp 1-8, 1996.

\item M.A. Nielsen and I.L. Chuang : Quantum Computation and Quantum Information, Cambridge University Press, Cambridge 1999.

\item I.R. Shafarevich, Basic Algebraic Geometry I, Second edition, Springer Verlag, Berlin. (Translated from Russian) 1994.

\end{enumerate}

\end{document}